\documentclass[11pt]{article}

\usepackage[utf8]{inputenc}
\usepackage{amsmath, amssymb, amsthm}
\usepackage[letterpaper, margin=1in]{geometry}
\usepackage{booktabs}
\usepackage{graphicx}
\usepackage{titlesec}

\usepackage{setspace}
\usepackage[authoryear, round]{natbib} 

\usepackage[colorlinks=true, citecolor=blue, linkcolor=blue, urlcolor=blue]{hyperref}

\title{\textbf{Fast Robust Regression via Orthogonal Block Updates \vspace{0.75cm}}}

\author{
	\Large \textbf{Anthony Christidis$^*$} \\
	Department of Biomedical Informatics, Harvard Medical School \\
	$^*$Corresponding Author: \href{mailto:anthony-alexander_christidis@hms.harvard.edu}{\texttt{anthony-alexander\_christidis@hms.harvard.edu}} \\[1.5em]
	\Large \textbf{Matías Salibián-Barrera} \\
	Department of Statistics, University of British Columbia
}

\date{\today} 

\begin{document}
	
	\maketitle
	
	\begin{abstract}
		Robust regression methods, particularly MM-estimators, are essential for analyzing datasets where heavy-tailed noise or high-leverage outliers may be present. Algorithms to compute these estimators are iterative and rely on having a good initial point. A widely-used probabilistic approach to obtaining an initial regression estimator that is not affected by outliers consists of fitting linear regression models to many random subsets of the training data. Unfortunately, the number of subsets that need to be considered in order to have a high probability of finding one that is clean of outliers grows exponentially with the number of predictors. This renders the approach unfeasible for models with a moderate to large numbers of variables. Alternative non-stochastic strategies that have been proposed recently also fail to scale well when the number of parameters and
		observations are large. To overcome this problem, we propose a highly scalable algorithm based on block-coordinate descent. Our Robust Orthogonal Block Updates (ROBU) algorithm uses lower-dimensional blocks of explanatory variables, which require much fewer sub-samples to find a good initial point with high probability. Extensive simulations and a proteogenomics data application illustrate the robustness properties of the estimators computed with ROBU and demonstrate that they compare favourably to those calculated with existing algorithms.
	\end{abstract}
	
	
	\newpage
	
	\section{Introduction} \label{sec:intro} 
	
	Robust regression methods are essential for analyzing datasets that may contain atypical observations, heavy-tailed noise, or high-leverage points. It is well-known that classical ordinary least squares (OLS) estimates can be highly sensitive to small such deviations from standard distributional assumptions \citep{maronna2019robust}. Among the most widely adopted robust regression estimators is the MM-estimator \citep{yohai1987high}, which simultaneously achieves maximal outlier protection and high statistical efficiency when no outliers are present in the training set. We refer the reader to \citet{maronna2019robust} for a comprehensive discussion. 
	
	MM-estimators are computed using re-weighted least squares iterations 
	(IRWLS) that start from a robust initial estimate, typically an S-estimator \citep{rousseeuw1984robust}. This initial S-estimate is the minimizer of an implicitly defined non-convex objective function that can also be computed using  an IRWLS algorithm \citep{salibian2006fast}. The robustness of the estimator depends on the initial point not being affected by potential outliers. The most common strategy to find a good initial point involves calculating a large number of random data-dependent starting values and choosing the resulting concentration point with the best value of the objective function. These initial points are obtained by fitting a linear model to small random subsets of the training data. In settings with $p$ explanatory variables, we randomly draw sub-samples of $p$ points. The idea is that most outlier-free random sub-samples are likely to yield a good starting point. If the training set contains $\epsilon \times 100 \%$ of outliers, then the probability of drawing one ``clean'' sub-sample of size $p$ is approximately $(1-\epsilon)^p$. It follows that the number of random sub-samples $N$ needed to find at least one good starting point with high probability increases exponentially with $p$ \citep{salibian2006fast}. 
	When the computation budget is not sufficient to generate enough random subsets these algorithms may fail to find a robust solution to the estimating equations, and return a biased estimate instead. This may occur when $p$ is even moderately large, as is the case in modern datasets in fields such as genomics, chemometrics, and signal processing. 
	
	A recently proposed partial solution to this problem 
	uses a fixed number of non-stochastic initial points \citep{hubert2012deterministic, maronna2019robust}. However, while these deterministic algorithms use a smaller number of candidate starting points, they  scale poorly as the dimension $p$ and the number of observations increase, rendering them equally impractical for high-dimensional applications.
	
	To overcome these challenges in the large $p$ regime (where $p < n$), we propose the Robust Orthogonal Block Updates (ROBU) algorithm. ROBU leverages the principles of block-coordinate descent \citep{tseng01} to partition the high-dimensional robust regression problem into a series of lower-dimensional ones. Block-coordinate descent methods can converge slowly on ill-conditioned problems, a situation that often arises when predictors are highly correlated; ROBU therefore works with an orthogonal basis of the design space, which improves the stability of the block updates \citep{nesterov2012efficiency, wright2015coordinate}.
	Furthermore, the proportion of outliers and their leverage remain unchanged in the orthogonal representation. Because the subsequent block-wise MM-estimators possess a 50\% breakdown point, they safely absorb and trim these rotated outliers. In addition, by controlling the dimension of each block, ROBU decreases the cost of the corresponding matrix operations, and, more importantly, also significantly reduces the number of random sub-samples needed to find at least one good initial point. In this way, we can retain a high probability of finding a reliable starting point at a fraction of the computational cost of the original problem. 
	
	
	The remainder of this paper is organized as follows. Section~\ref{sec:methodology} presents the ROBU algorithm, while Section~\ref{sec:theory} discusses the properties of the algorithm, including the reduction in complexity it can achieve. Section~\ref{sec:sims} presents an extensive simulation study benchmarking ROBU against standard robust methods, alongside a numerical sensitivity analysis for the block dimension hyperparameter. An illustrative proteogenomics application is included in Section~\ref{sec:application} and Section~\ref{sec:conclusion} provides concluding remarks.
	
	\section{Methodology: The ROBU Algorithm} \label{sec:methodology}
	
	Consider $n$ points $(y_i, \mathbf{x}_i^\top)^\top \in \mathbb{R}^{p+1}$, 
	$i=1, \ldots, n$, assumed to follow a linear regression model 
	$$
	y_i = \mathbf{x}_i^\top \boldsymbol{\beta}_0 + e_i \, , \quad i = 1, \dots, n \, ,
	$$
	where $y_i \in \mathbb{R}$ is the response variable, $\mathbf{x}_i \in \mathbb{R}^p$ is a vector of predictors, $\boldsymbol{\beta}_0 \in \mathbb{R}^p$ is the vector of unknown regression coefficients, and $e_i$ are independent errors, symmetric around zero and independent from the $\mathbf{x}_i$'s. 
	In matrix notation: $\mathbf{y} = \mathbf{X} \, \boldsymbol{\beta}_0 + \mathbf{e}$, 
	where $\mathbf{y} \in \mathbb{R}^n$, $\mathbf{X} \in \mathbb{R}^{n \times p}$,
	and $\mathbf{e} \in \mathbb{R}^n$. 
	
	We are concerned with the practical situation where the training set may contain outliers. The highly robust MM-estimator \citep{yohai1987high} is computed in two stages. First, let $\boldsymbol{\hat{\beta}}_S$ be an initial high-breakdown S-estimator \citep{rousseeuw1984robust} satisfying
	\begin{equation} \label{eq:s_est}
		\boldsymbol{\hat{\beta}}_S = \arg \min_{\boldsymbol{\beta} \in \mathbb{R}^p} \, s(\boldsymbol{\beta}) \, ,
	\end{equation}
	where $s(\boldsymbol{\beta})$ satisfies 
	$$
	\frac{1}{n} \sum_{i=1}^n \rho_0 \left( \frac{y_i - \mathbf{x}_i^\top \boldsymbol{\beta}}{s(\boldsymbol{\beta})} \right) = \delta \, ,
	$$
	and where $\rho_0 : \mathbb{R} \to [0, +\infty)$ is bounded, symmetric, and non-decreasing on $(0, +\infty)$. A commonly used class of such functions 
	is given by Tukey's bi-square family \citep{maronna2019robust}. To ensure
	consistency of these estimators, the tuning parameter 
	$\delta \in (0, 0.5]$ satisfies $E[ \rho_0( e) ] = \delta$. The associated residual M-scale 
	estimator is
	\begin{equation} \label{eq:s-scale}
		\hat{\sigma}_S = s(\boldsymbol{\hat{\beta}}_S) \, .
	\end{equation}
	
	To increase the statistical efficiency of the regression estimator, the S-estimator is used as the initial value for iterative re-weighted least 
	squares (IRWLS) iterations    
	to compute an M-estimate $\boldsymbol{\hat{\beta}}_M$
	\begin{equation} \label{eq:m_est}
		\boldsymbol{\hat{\beta}}_M = \arg\min_{\boldsymbol{\beta} \in \mathbb{R}^p} \ \frac{1}{n} \, 
		\sum_{i=1}^n \rho_1 \left( \frac{y_i - \mathbf{x}_i^\top \boldsymbol{\beta}}{\hat{\sigma}_S} \right) \, ,
	\end{equation}
	where the residual scale estimator $\hat{\sigma}_S$ is as in \eqref{eq:s-scale}, 
	$\rho_1(u) \le \rho_0(u)$, $u \in \mathbb{R}$, and $\rho_1$ satisfies the same regularity conditions as $\rho_0$ \citep{maronna2019robust}.
	The IRWLS algorithm is based on the first order conditions for the 
	minimum above:
	\[
	\frac{1}{n} \sum_{i=1}^n \rho_1' \left( \frac{y_i - \mathbf{x}_i^\top \hat{\boldsymbol{\beta}}_M}{\hat{\sigma}_S} \right) \, \mathbf{x}_i \, = \, 
	\mathbf{0} \, ,
	\]
	where $\rho_1'$ is the derivative of $\rho_1$. The above equation
	can be re-written as 
	\begin{equation} \label{eq:w_irwls}
		\left[ \sum_{i=1}^n w_i( \boldsymbol{\hat{{\beta}}}_M ) \, \mathbf{x}_i \, 
		\mathbf{x}_i^\top \right] \, \boldsymbol{\hat{\mathbf{\beta}}}_M \, = \, 
		\sum_{i=1}^n w_i( \boldsymbol{\hat{{\beta}}}_M ) \, \mathbf{x}_i \, y_i \, ,
	\end{equation}
	where $w_i( \boldsymbol{\hat{{\beta}}}_M )  = 
	\rho_1' \left( \tilde{r}_i \right) / \tilde{r}_i$ and 
	$\tilde{r}_i = (y_i - \mathbf{x}_i^\top \boldsymbol{\hat{{\beta}}}_M
	) / \hat{\sigma}_S$.
	Given a current vector of regression parameters $\boldsymbol{\hat{\beta}}^{(j)}$, 
	\eqref{eq:w_irwls} suggests using the following weighted least
	squares iterations:
	\begin{equation} \label{eq:iters}
		\boldsymbol{\hat{{\beta}}}^{(j+1)}_M \, = \, \left[ \sum_{i=1}^n w_i( \boldsymbol{\hat{{\beta}}}_M^{(j)} ) \, \mathbf{x}_i \, 
		\mathbf{x}_i^\top \right]^{-1} \, 
		\left[ \sum_{i=1}^n w_i( \boldsymbol{\hat{{\beta}}}_M^{(j)} ) \, 
		\mathbf{x}_i \, y_i \right] \, , \quad j = 0, 1, \ldots \, .
	\end{equation}
	Similar IRWLS iterations are useful to 
	compute the S-estimator in \eqref{eq:s_est}
	\citep{salibian2006fast}. Under standard regularity conditions
	these iterations decrease the objective function \citep{salibian2006fast}.
	Commonly used convergence criteria look at consecutive
	$\boldsymbol{\hat{{\beta}}}^{(j)}_M$'s, or at the value of the 
	objective function. 
	
	At the last iteration of \eqref{eq:iters} 
	the weights can be used to detect outliers in the training set. 
	Specifically, we expect
	atypical observations to have large
	residuals with respect to the robust regression fit, 
	e.g. $|\tilde{r}_i| > 3$. Since the
	function $\rho_1$ is bounded (generally constant) for large values of its
	argument, outlying points will have   
	$w_i(\boldsymbol{\hat{{\beta}}}_M ) = \rho_1'(\tilde{r}_i) / \tilde{r}_i \approx 0$.
	
	The success of iterative algorithms such as IRWLS relies on using a 
	large number of random initial values,
	and choosing the concentration point with the best  
	objective function. To compute robust estimators, the random 
	starting points are generally constructed by 
	fitting a linear model to random subsets of the training dataset. 
	The intuition behind this strategy is that most outlier-free 
	sub-samples will provide an initial point that represents 
	well the ``good'' observations, and hence results in a good 
	fit for the non-outlying points. 
	To ensure a high probability of finding a ``clean'' starting point, the number
	of sub-samples has to increase
	exponentially with $p$ (see the discussion in Section 
	\ref{sec:theory}). When $p$ is large (e.g., hundreds of variables), this search can become computationally prohibitive, and the algorithm may return a biased estimate affected by the outliers. 
	
	\subsection{The ROBU Algorithm} \label{subsec:algo}
	
	To address these challenges we propose the Robust Orthogonal Block Updates (ROBU) algorithm for MM-estimators. ROBU leverages a divide-and-conquer strategy, partitioning the high-dimensional problem into lower-dimensional ones, which can be thought of as one pass of block-coordinate descent. The fits for these lower-dimensional problems require exploring different starting points as discussed above. However, by controlling the number of variables in
	each of these subproblems, we can achieve an overall high-probability of success with a significantly lower computational budget than the original problem. 
	
	Block-coordinate descent methods can converge slowly on ill-conditioned problems, a situation that often arises when explanatory variables are highly correlated \citep{nesterov2012efficiency, wright2015coordinate}. A natural solution is to ``orthogonalize'' the 
	features using the QR decomposition of the design matrix
	\citep{golub1996matrix}: 
	\begin{equation} \label{eq:qr}
		\mathbf{X} \, = \, \mathbf{Q} \, \mathbf{R} \, ,
	\end{equation}
	where $\mathbf{Q} \in \mathbb{R}^{n \times p}$ has orthonormal columns ($\mathbf{Q}^\top \, \mathbf{Q} = \mathbf{I}_{p \times p}$), and $\mathbf{R} \in \mathbb{R}^{p \times p}$ is upper-triangular. The regression model can then be  re-parameterized as:
	\begin{equation} \label{eq:q-model}
		\mathbf{y} \, = \, \mathbf{X} \, \boldsymbol{\beta}_0 + \mathbf{e} \, = \, 
		\mathbf{Q} \boldsymbol{\theta}_0 + \mathbf{e} \, , \quad \text{where} \quad \boldsymbol{\theta}_0 = \mathbf{R} \, \boldsymbol{\beta}_0 \, .
	\end{equation}
	Since the features in the ``Q'' design matrix are linearly uncorrelated, we expect the block-coordinate updates to proceed efficiently.
	
	The main parameter of the ROBU algorithm is the number of blocks 
	$1 \le k \le p$ to be used. Given $k$, let $\mathcal{I}_1, \dots, \mathcal{I}_k$ be a partition of the set $\{1,\dots,p\}$ into $k$ (approximately equally
	sized) disjoint subsets, and let 
	$\mathbf{Q}_j \in \mathbb{R}^{n \times p_j}$, 
	$1 \le j \le k$, be the corresponding sub-matrices of $\mathbf{Q}$ formed by the columns indexed by each $\mathcal{I}_j$, where $p_j = |\mathcal{I}_j|$ and 
	$\sum_{j=1}^k p_j = p$. The goal is to obtain a robust estimator for $\boldsymbol{\theta}_0$ in \eqref{eq:q-model}, which can then be transformed back into an estimate for $\boldsymbol{\beta}_0$. 
	
	The ROBU algorithm is as follows. 
	
	\medskip
	\noindent\textbf{The ROBU Algorithm:}
	\begin{enumerate}
		\item \textbf{Orthogonal decoupling.} Compute the QR decomposition 
		$\mathbf{X} = \mathbf{Q} \, \mathbf{R}$.
		\item \textbf{Block partitioning.} Choose an integer $1 \le k \le p$, and construct a partition $\mathcal{I}_1, \dots, \mathcal{I}_k$ of $\{1,\dots,p\}$. Define $\mathbf{Q}_j$ as above for $j = 1,\dots,k$.
		\item \textbf{Blockwise S-initialization.} Initialize the working residuals $\mathbf{r}^{(0)} = \mathbf{y}$ and set $\boldsymbol{\hat{\theta}}_{\mathrm{init}} = \mathbf{0} \in \mathbb{R}^p$. For $j = 1,\dots,k$:
		\begin{enumerate}
			\item Compute an S-estimate $\boldsymbol{\hat{\theta}}_j \in \mathbb{R}^{p_j}$ for the regression of $\mathbf{r}^{(j-1)}$ on $\mathbf{Q}_j$, as in Equation~\eqref{eq:s_est} and a subsampling-based algorithm such as Fast-S \citep{salibian2006fast}.
			\item Update the residuals as $\mathbf{r}^{(j)} = \mathbf{r}^{(j-1)} - 
			\mathbf{Q}_j \hat{\theta}_j$.
			\item Insert $\boldsymbol{\hat{\theta}}_j$ into the corresponding coordinates of $\boldsymbol{\hat{\theta}}_{\mathrm{init}}$ (those indexed by $\mathcal{I}_j$).
		\end{enumerate}
		\item \textbf{Global S-estimate.} Using the $\boldsymbol{\hat{\theta}}_{\mathrm{init}}$ obtained at the end of the loop above as a starting value, compute the S-estimate $\boldsymbol{\hat{\theta}}_S$ for the model 
		$\mathbf{y} = \mathbf{Q} \boldsymbol{\theta} + \mathbf{e}$ as in Equation~\eqref{eq:s_est} via the usual iteratively reweighted least squares (IRWLS) scheme; see \citet{maronna2019robust} for details.
		\item \textbf{Final M-estimate.} Starting from $\boldsymbol{\hat{\theta}}_S$ and keeping the scale fixed at $\hat{\sigma}_S$, compute the final M-estimate $\boldsymbol{\hat{\theta}}_M$ for the orthogonal model by minimizing the M-objective in Equation~\eqref{eq:m_est}, using standard IRWLS iterations.
		\item \textbf{Back-transformation.} Obtain the robust regression coefficients in the original predictor space $\boldsymbol{\hat{\beta}}_M$ by solving the upper-triangular system $\mathbf{R} \, \boldsymbol{\hat{\beta}}_M = \boldsymbol{\hat{\theta}}_M$. 
	\end{enumerate}
	
	The proposed ROBU algorithm consists of three core steps. First, the QR decomposition results in uncorrelated predictors, so that fitting on successive blocks $\mathbf{Q}_j$ using the current residuals $\mathbf{r}^{(j-1)}$ results in stable partial regression coefficients, without the blocks competing to explain the same variation in the response variable. Second, by bounding the block size $p_j$ (discussed in Section~\ref{subsec:blocks}), the blockwise S-estimates in Step~3 operate in a regime where the probability of drawing at least one clean elemental sub-sample can be kept high with notably fewer random attempts than when solving the full $p$-dimensional problem. Finally, Steps~4--6 use this robust initialization with the usual MM-estimator algorithm and back-transforms it through $\mathbf{R}$.
	
	\subsection{Selection of the Number of Blocks ($k$)} \label{subsec:blocks}
	
	The number of blocks $k$ in ROBU determines the  trade-off
	between computational cost and robustness of the estimator. Small values  
	of $k$ result in sub-problems with larger numbers of variables, 
	which require an exponentially larger number of random sub-samples to
	find a ``clean'' initial point with high probability. By increasing $k$ 
	(and thus dividing the predictor space into lower-dimensional blocks), ROBU only
	requires $k$ sets of remarkably fewer random subsets to find a good initial point. 
	At the same time, finding these candidate initial points involve inverting lower-dimensional matrices hence ROBU can achieve an important reduction in overall computational cost.
	However, large values of $k$ (e.g., $k \approx p$) result in blocks with very few variables, and the block updates may fail to detect multivariate high-leverage outliers, which may remain masked in low-dimensional marginal distributions
	\citep{rousseeuw90}. Our experiments suggest that using blocks with 10 to 20 variables provides an optimal balance: it ensures a high probability of outlier-free sub-sampling while preserving local multivariate signal. Therefore, we set the default number of blocks to $k = \max(1, \lfloor p/10 \rfloor)$. An empirical sensitivity analysis evaluating the impact of different choices of $k$ is presented in Section~\ref{subsec:sensitivity}.
	
	\section{Theoretical and Computational Properties} \label{sec:theory}
	
	In this section we study the properties of the estimators computed 
	with the ROBU algorithm. In particular, we quantify the substantial reduction in computational cost needed to find a ``clean'' initial point with high probability, and also show that the proportion of outliers and their
	corresponding leverage measures in the QR representation of the design matrix are the same as in the original variables. %
	These observations are critical for the reliable application of 
	ROBU to contaminated datasets. 
	
	\subsection{Computational Cost Reduction of ROBU}
	
	As discussed in Section \ref{sec:methodology}, the IRWLS algorithms used to compute $\boldsymbol{\hat{\beta}}_S$ and $\boldsymbol{\hat{\beta}}_M$ require a reliable initial point. The usual strategy for doing this considers a large number of candidates 
	obtained from linear model fits to randomly chosen subsets of the data. 
	Given the proportion of outliers in the training set we can calculate the relationship between the number of random subsets 
	under consideration and the probability of at least one of them not  
	including any outliers. Specifically, 
	if $\epsilon \in [0, 1/2)$ is the proportion of outliers in the data set, $p$ is the number of explanatory variables, and $N_s$ is the number of subsets of size $p$ randomly chosen from the $n$ available observations, then the probability of finding at least one clean subset is approximately
	\begin{equation} \label{eq:prob-clean}
		P_{s} = 1 - \bigl\{1 - (1-\epsilon)^p\bigr\}^{N_s} \, ,
	\end{equation}
	\citep{rousseeuw-84, salibian2006fast}.
	The same reasoning shows that the corresponding success probability when 
	applying this construction to the $k$ blocks with $N_{robu}$ subsets 
	equally distributed across blocks is approximately
	\begin{equation} \label{eq:prob-robu}
		P_{robu} = \left[ 1 - \bigl\{1 - (1-\epsilon)^{p/k}\bigr\}^{N_{robu}/k} \right]^k.
	\end{equation}
	An alternative way of presenting these results is to calculate the 
	smallest number of 
	random sub-samples needed in order to find a good initial point with probability at least $1
	- \alpha$, for a fixed $\alpha \in (0, 1)$. It is easy to see that they are 
	$$
	N_s = \frac{ \log(\alpha) }{ \log(1-(1-\epsilon)^p)} \, 
	\approx \, \frac{ \log(\alpha) }{ -(1-\epsilon)^p } \, ,
	$$
	and 
	$$
	N_{robu} = \frac{ k \log( 1 - (1- \alpha)^{1/k} )}{ 
		\log(1 - (1-\epsilon)^{p/k})} \, , 
	$$
	for the standard and ROBU algorithms, respectively.  
	To illustrate the magnitude of the difference between $N_s$ and $N_{robu}$, consider a dataset with $p=200$ 
	explanatory variables, $\epsilon=0.15$ contamination, and a goal of 99\% probability
	of success. The standard estimator will need to evaluate $N_s \approx 6 \times 10^{14}$ sub-samples, which is computationally impossible in practice. However, if we partition the space into $k=20$ blocks of $p/k = 10$ variables each, ROBU achieves the same 99\% probability of success using a total budget of $N_{robu} \approx 693$ sub-samples (35 sub-samples per block). 
	
	Another factor that affects the computational complexity of finding the  
	initial regression estimator is the cost of calculating  
	the random candidates. For the standard subsampling algorithm,
	each of these candidates is calculated by solving
	a $p \times p$ linear system of equations, so the total cost 
	is $\mathcal{O}(N_s \, p^3)$. 
	For ROBU, the size of each of these linear systems is $(p/k) \times (p/k)$, 
	but we also need to include the cost of the QR decomposition,
	which is $\mathcal{O}(n \, p^2)$ \citep{golub1996matrix}. Thus, the 
	total cost of exploring $N_{robu}$ candidates with ROBU is 
	$\mathcal{O}(N_s \, p^3 / k^3 + n \, p^2)$. The bulk of the 
	computational complexity 
	advantage for ROBU is a result of its required number of candidates
	$N_{robu}$ is typically several orders of magnitude smaller than $N_s$. 
	
	
	A similar computational bottleneck affects recently proposed 
	deterministic algorithms to
	compute robust estimators. The starting values for deterministic MM-estimators are typically derived using the Peña-Yohai (PY) algorithm \citep{penayohai99}. The complexity of this algorithm
	is difficult to assess since it iterates the computation of M-scale estimators. But for each of these iterations one has to solve $\mathcal{O}(n)$ LS estimators, sort $n$ residuals, compute the diagonal of the ``hat'' matrix, and the eigenvalues of a $p \times p$ matrix (which involves calculating a square root of another $p \times p$ matrix). Each iteration has a
	computational complexity of $\mathcal{O}(p \, n \log(n) + n \, p^2 
	+ p^3)$ \citep{golub1996matrix, eigencomplex}.
	
	
	As an example, consider a training set with $n = \text{10,000}$ observations, 
	$p = 500$ explanatory variables, and $\epsilon = 0.15$ (15\% outliers). 
	If we set the minimum probability of success at 99\% ($\alpha = 0.01$), then 
	the standard algorithm would require $N_s \, p^3 \approx 9 \times 10^{35}$ 
	operations, the deterministic alternative needs to iterate a loop that requires 
	$p \, n \log(n) + n \, p^2 
	+ p^3 \approx 3 \times 10^9$ operations in each pass. 
	Using $k = 50$ blocks, ROBU only needs one set of 
	$N_s \, p^3 / k^3 + n \, p^2 \approx 3 \times 10^9$ operations. 
	Our experiments reported in Section \ref{sec:sims} below 
	illustrate the magnitude of the timing advantage of ROBU over the deterministic
	algorithm. In practice ROBU requires a fraction of the time 
	used by standard methods, while maintaining equal or better accuracy.  
	
	
	\subsection{Invariance of Outliers Structure Over QR Transformation} \label{sec:invariance}
	
	When the design matrix $\mathbf{X}$ contains
	outliers, it would not be desirable that the QR-transformed
	explanatory variables $\mathbf{q}_i = \mathbf{R}^{-\top} \mathbf{x}_i$, 
	$1 \le i \le n$, contained 
	more (or more extreme) outliers than those in the original training set. 
	In this Section we show that this is not the case.
	
	Outliers in a linear regression context can be either ``vertical'' or
	``high-leverage''. Vertical outliers deviate from the bulk of the data
	only in the values of their response variable. 
	High-leverage outliers are atypical  
	in both their response and explanatory variables simultaneously, 
	and are the most damaging and difficult to control. 
	To show that vertical
	outliers in the transformed variables coincide with those 
	in the original data set note that the QR transformation \eqref{eq:qr} 
	implies the 
	transformation $\boldsymbol{\beta} \to 
	\boldsymbol{\theta} = \mathbf{R} \, \boldsymbol{\beta}$
	in the vector of regression coefficients. Here we assume
	that the design matrix 
	$\mathbf{X}$ is full rank, so that 
	$\mathbf{R}^{-\top} = (\mathbf{R}^{-1})^\top = (\mathbf{R}^{\top})^{-1}$.
	Hence, computing the S- and MM-estimators on the transformed explanatory 
	variables $\mathbf{q}_i$'s is the same as 
	mapping the estimators computed with the original $\mathbf{x}_i$'s. In symbols, 
	$\boldsymbol{\hat{\theta}}_M = \mathbf{R} \, \boldsymbol{\hat{\beta}}_M$, and similarly for
	the S-estimator. It follows that the associated residual scale 
	estimator \eqref{eq:s-scale} remains invariant as well, and 
	\[
	\frac{y_i - \mathbf{x}_i^\top \boldsymbol{\hat{\beta}}_M}{\hat{\sigma}_S} = 
	\frac{y_i - \mathbf{q}_i^\top \boldsymbol{\hat{\theta}}_M}{\hat{\sigma}_S} \, .
	\]
	This also implies  
	$w_i( \boldsymbol{\hat{\beta}}_M ) = w_i( \boldsymbol{\hat{\theta}}_M)$, $i=1, \ldots, n$.
	
	Furthermore, high-leverage outliers are also identical in both 
	basis representations. To see this, recall that the leverage 
	measures the distance of each point to 
	the center of its reference distribution. The most commonly used leverage 
	measure is the Mahalanobis distance
	\[
	d_{X, i}^2 \, = \, \left( \mathbf{x}_i - \boldsymbol{\mu}_X \right)^\top \, 
	\mathbf{\Sigma}_X^{-1} \, 
	\left( \mathbf{x}_i - \boldsymbol{\mu}_X \right) \, ,
	\]
	where $\boldsymbol{\mu}_X \in \mathbb{R}^p$ and $\boldsymbol{\Sigma}_X \in \mathbb{R}^{p \times p}$ are the
	location and scatter parameters of the distribution of the explanatory
	variables $\mathbf{x}_i$. A natural property 
	of multivariate
	location and scatter parameters is that they are
	affine equivariant, which means 
	that if $\mathbf{A} \in \mathbb{R}^{k \times p}$ and 
	$\mathbf{b} \in \mathbb{R}^k$ and the data are transformed 
	$\mathbf{x}_i \to \mathbf{z}_i = \mathbf{A} \mathbf{x}_i 
	+ \mathbf{b}$, then the corresponding location and scatter parameters
	change accordingly: 
	$\boldsymbol{\mu}_Z = \mathbf{A} \, \boldsymbol{\mu}_X + \mathbf{b}$ and
	$\boldsymbol{\Sigma}_Z = \mathbf{A} \, \boldsymbol{\Sigma}_X 
	\mathbf{A}^\top$, respectively. Thus, it is easy to check that 
	\[
	\left( \mathbf{x}_i - \boldsymbol{\mu}_X \right)^\top \, 
	\mathbf{\Sigma}_X^{-1} \, 
	\left( \mathbf{x}_i - \boldsymbol{\mu}_X \right) \, = \, 
	\left( \mathbf{q}_i - \boldsymbol{\mu}_Q \right)^\top \, 
	\mathbf{\Sigma}_Q^{-1} \, 
	\left( \mathbf{q}_i - \boldsymbol{\mu}_Q \right) \, .
	\]
	Consequently, the leverage of each observation $\mathbf{x}_i$ relative to the rest of the sample is preserved for the corresponding $\mathbf{q}_i$ in the orthogonal space.
	

	\section{Simulation Study} \label{sec:sims}
	
	We report the results of a simulation study conducted to 
	evaluate the computational efficiency and statistical robustness of the proposed ROBU algorithm. The primary objective is to show that ROBU substantially reduces computation time in high dimensions while mitigating the probabilistic algorithmic failure associated with standard resampling techniques, and maintaining 
	the statistical properties of the estimator.
	
	\subsection{Simulation Design} \label{subsec:sim_design}
	
	To study the scaling behavior of the algorithms we 
	generated synthetic datasets with a fixed sample size of $n = 1{,}500$ and varied the number of predictors $p \in \{100, 200, 400\}$. To mimic 
	the multicollinearity frequently encountered in high-dimensional datasets (such as genomic pathways or spatial data), the predictors $\mathbf{x}_i \in \mathbb{R}^p$ were sampled from a multivariate normal distribution $\mathcal{N}(\mathbf{0}, \boldsymbol{\Sigma})$. The covariance matrix $\boldsymbol{\Sigma}$ had a block-diagonal structure consisting of 10 blocks. Within each block, the variables shared a pairwise correlation of $\rho = 0.5$, while variables in different blocks were uncorrelated; all marginal variances were equal to 1.
	
	The true regression coefficients $\boldsymbol{\beta} \in \mathbb{R}^p$ were drawn from a standard normal distribution $\mathcal{N}(\mathbf{0},\mathbf{I})$. The response variable $y_i$ follows the linear model $y_i = \mathbf{x}_i^T \boldsymbol{\beta} + e_i$, where the errors $e_i$ have a normal distribution with mean zero, and variance scaled for each simulated dataset to maintain a Signal-to-Noise Ratio (SNR) of 3. Finally, since the data-generating model has zero intercept, we fit all methods without an intercept term.
	
	\subsection{Contamination Scenarios} \label{subsec:sim_scenarios}
	To stress-test the estimators, we evaluated three data contamination scenarios across contamination proportions $\epsilon \in \{0.10, 0.20, 0.30\}$. For the contaminated settings, a fixed proportion $\epsilon$ of the $n$ observations were randomly selected and replaced according to the following mechanisms:
	\begin{enumerate}
	\item \textbf{Clean Data:} The baseline scenario with $\epsilon = 0$, where all $n$ observations are generated from the multivariate normal model described in Section~\ref{subsec:sim_design}.
	\item \textbf{Vertical Outliers:} The predictor matrix $\mathbf{X}$ remains unaltered. For the contaminated observations, the response variable $y_i$ was shifted away from the true hyperplane by adding a bias term of magnitude $200\,\sigma_e$, where $\sigma_e$ is the standard deviation of the error term $e_i$. This construction produces \emph{pure response} contamination, in which atypicality occurs only in the $y$-direction.
	\item \textbf{High-Leverage Outliers:} This scenario is designed to be challenging for subsampling-based initialization. For the contaminated observations, 10\% of the predictor coordinates (sampled uniformly at random to prevent block-alignment artifacts) were corrupted by adding a coordinate-wise shift drawn from a $\mathcal{N}(10,1)$ distribution. This creates sparse (coordinate-wise) contamination in the covariates, while keeping the remaining coordinates distributed as $\mathcal{N}(\mathbf{0},\boldsymbol{\Sigma})$. Simultaneously, the corresponding response variables $y_i$ were generated using a different coefficient vector $\boldsymbol{\beta}_{\text{bad}} \sim \mathcal{N}(-5,2^2)$, so that the contaminated observations form a coherent alternative regression relationship and induce a systematic bias in the fitted regression hyperplane.
	\end{enumerate}
	
	\subsection{Competitors and Implementation} \label{subsec:sim_competitors}
	We compared the performance of four estimators:
	\begin{itemize}
	\item \textbf{Standard OLS:} The classical Ordinary Least Squares estimator, serving as a baseline to illustrate the failure of non-robust methods under contamination.
	\item \textbf{Standard MM:} The MM-estimator \citep{yohai1987high} computed via a Fast-S initialization \citep{salibian2006fast}, implemented using the \texttt{lmrob} function from the \texttt{robustbase} package \citep{maechler2023robustbase}.
	\item \textbf{Deterministic MM:} A deterministic counterpart to the MM-estimator that avoids random subsampling for initialization \citep{hubert2012deterministic}, implemented via the \texttt{lmrobdetMM} function from the \texttt{RobStatTM} package \citep{maronna2019robust}.
	\item \textbf{ROBU:} Robust Orthogonal Block Updates (Section~\ref{subsec:algo}). 
	\end{itemize}
	
	To facilitate comparison, the Standard MM-estimator and ROBU were configured to use the
	same loss functions, nominal breakdown point (50\%), and convergence tolerances (via a common
	\texttt{lmrob.control} list). For the resampling-based procedures, we used \texttt{nResample} =
	2{,}000 for $p \in \{100, 200\}$ and \texttt{nResample} = 5{,}000 for $p = 400$, holding these budgets
	fixed across all contamination levels $\epsilon$. In ROBU, this same \texttt{nResample} value was applied
	within each block during initialization, so that ROBU typically explores more elemental subsets in
	total than the standard MM-estimator, but in lower-dimensional subproblems, leading to a significantly reduced
	wall-clock time. Appendix~\ref{sec:appendixA} reports an additional experiment in which the total
	resampling budget is distributed across blocks; in that setting, ROBU achieves similar
	estimation accuracy while decreasing computation time even further, 
	which is consistent with the analysis in
	Section~\ref{sec:theory}. In all configurations, the number of blocks was chosen according to the
	heuristic in Section~\ref{subsec:blocks}, $k = \max(1, \lfloor p/10 \rfloor)$, so that for
	$p \in \{100, 200, 400\}$ we used $k \in \{10, 20, 40\}$ blocks, respectively.
	
	Performance was evaluated across 50 independent replications for each simulation configuration. Estimation accuracy was quantified using the mean squared estimation error
	\[
	\text{MSE} = \frac{1}{50} \, \sum_{j=1}^{50} \left\lVert\boldsymbol{\hat{\beta}}^{(j)} - \boldsymbol{\beta}_{\text{true}}\right\rVert_2^2 \, .
	\]
	Furthermore, to evaluate robust detection accuracy, we tracked the average number of True Positives (TP) and False Positives (FP) across replications. An observation was formally classified as a trimmed outlier if its final robust weight $w_i(\boldsymbol{\hat{\beta}}_M)$ in \eqref{eq:w_irwls} was zero, providing a strict measure of how effectively each algorithm isolated the contamination.
	
	\subsection{Simulation Results} \label{subsec:sim_results}
	
\begin{figure}[h!]
	\centering
	\includegraphics[width=0.95\textwidth]{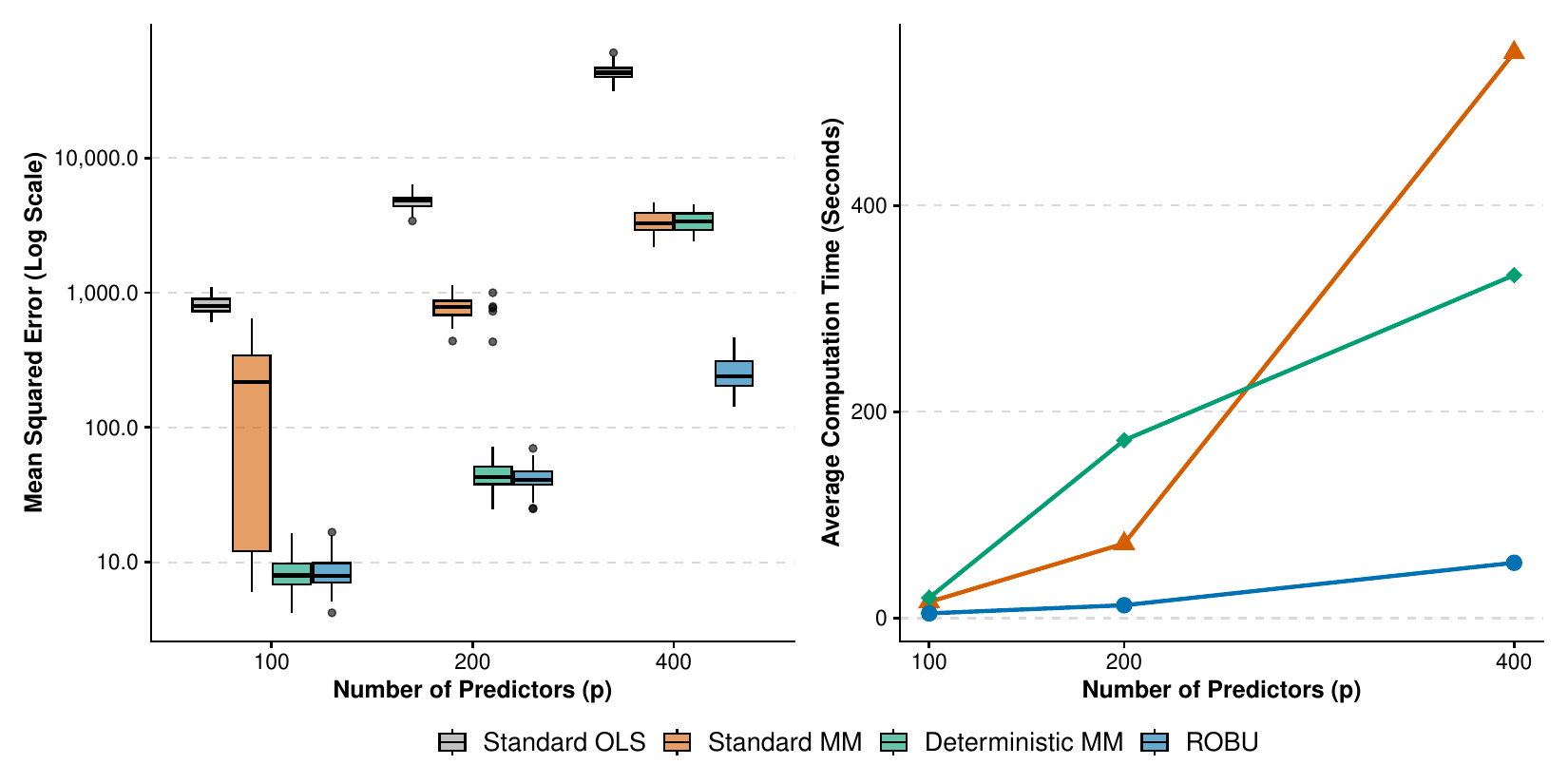}
	\caption{Performance of the estimators under high-leverage outliers (fixed at $\epsilon=0.20$, $n=1{,}500$). \textbf{(Left)} Boxplots of estimation accuracy (MSE) across 50 replications on a logarithmic scale. \textbf{(Right)} Average computation time (Seconds) across 50 replications.}
	\label{fig:main_plots}
	\end{figure}

	The simulation results are summarized in Table~\ref{tab:main_results}. To illustrate computational
	scaling and the effects of subsampling-based initialization under contamination,
	Figure~\ref{fig:main_plots} isolates performance under a fixed 20\% contamination regime. Extended
	tabular results for Scenario~2 (Vertical Outliers) are provided in Appendix~\ref{sec:appendixB}.
	
	\begin{table}[ht]
	\centering
	\caption{Average Mean Squared Error (MSE) and Computation Time (Seconds) under baseline Clean Data ($\epsilon=0$) and Scenario~3 (High-Leverage Outliers) across 50 replications ($n=1{,}500$). \vspace{0.5em}}
	\label{tab:main_results}
	
	\renewcommand{\arraystretch}{1.2}
	\setlength{\tabcolsep}{4pt}
	\begin{tabular}{ll c rr c rr c rr c rr}
		\toprule
		& & & \multicolumn{2}{c}{\textbf{Clean Data}} & & \multicolumn{8}{c}{\textbf{High-Leverage Outliers}} \\
		\cmidrule(lr){4-5} \cmidrule(lr){7-14}
		& & & \multicolumn{2}{c}{$\boldsymbol{\epsilon = 0}$} & & \multicolumn{2}{c}{$\boldsymbol{\epsilon = 0.10}$} & & \multicolumn{2}{c}{$\boldsymbol{\epsilon = 0.20}$} & & \multicolumn{2}{c}{$\boldsymbol{\epsilon = 0.30}$} \\
		\cmidrule(lr){1-2} \cmidrule(lr){4-5} \cmidrule(lr){7-8} \cmidrule(lr){10-11} \cmidrule(lr){13-14}
		$\boldsymbol{p}$ & \textbf{Method} & & \multicolumn{1}{c}{\textbf{MSE}} & \multicolumn{1}{c}{\textbf{Time}} & & \multicolumn{1}{c}{\textbf{MSE}} & \multicolumn{1}{c}{\textbf{Time}} & & \multicolumn{1}{c}{\textbf{MSE}} & \multicolumn{1}{c}{\textbf{Time}} & & \multicolumn{1}{c}{\textbf{MSE}} & \multicolumn{1}{c}{\textbf{Time}} \\
		\cmidrule(lr){1-2} \cmidrule(lr){4-5} \cmidrule(lr){7-8} \cmidrule(lr){10-11} \cmidrule(lr){13-14}
		
		100 & OLS     & & 4.4   & 0.0   & & 484.0   & 0.0   & & 649.4   & 0.0   & & 816.3   & 0.0 \\
		& Std MM  & & 4.9   & 42.3  & & 13.7    & 37.3  & & 102.8   & 37.5  & & 233.3   & 38.8 \\
		& Det MM  & & 4.8   & 55.9  & & 5.1     & 53.5  & & 5.8     & 54.7  & & 6.6     & 53.6 \\
		& ROBU    & & 4.7   & 12.0  & & 5.1     & 11.6  & & 5.9     & 10.5  & & 6.7     & 9.0 \\
		
		\addlinespace[0.7em]
		
		200 & OLS     & & 19.5  & 0.1   & & 2,097.1 & 0.1   & & 3,122.1 & 0.1   & & 4,193.1 & 0.1 \\
		& Std MM  & & 21.2  & 194.4 & & 92.9    & 201.2 & & 592.3   & 198.1 & & 828.7   & 178.6 \\
		& Det MM  & & 22.0  & 407.0 & & 21.9    & 404.3 & & 26.2    & 404.7 & & 127.1   & 419.1 \\
		& ROBU    & & 21.1  & 32.0  & & 21.9    & 31.4  & & 26.3    & 30.7  & & 30.7    & 21.4 \\
		
		\addlinespace[0.7em]
		
		400 & OLS     & & 92.5  & 0.1   & & 14,645.7 & 0.1  & & 25,175.6 & 0.1  & & 32,043.2 & 0.1 \\
		& Std MM  & & 108.6 & 885.8 & & 1,301.4  & 904.0 & & 1,864.9  & 891.1 & & 3,591.5  & 878.5 \\
		& Det MM  & & 114.2 & 1,893.7& & 120.1    & 1,987.2& & 585.3    & 2,079.1& & 3,687.0  & 2,335.0 \\
		& ROBU    & & 108.3 & 90.5  & & 118.0    & 86.5  & & 122.7    & 85.1  & & 160.5    & 68.0 \\
		\bottomrule
	\end{tabular}
	\end{table}
	
	Under clean data ($\epsilon=0$), all robust estimators achieve similar MSEs across dimensions,
	indicating that ROBU's blockwise initialization does not degrade estimation accuracy in the absence
	of outliers. Their computational profiles differ substantially. As shown in the right panel of
	Figure~\ref{fig:main_plots}, the resampling-based MM implementation becomes much slower as $p$
	increases (e.g., $\approx 886$ seconds per fit at $p=400$), consistent with the increasing cost of
	repeated dense linear-algebra operations. The deterministic baseline is even more expensive in this
	regime (e.g., $\approx 1{,}894$ seconds at $p=400$). By fitting lower-dimensional subproblems during
	initialization, ROBU reduces computation time, with runtimes on the order of $\approx 90$ seconds
	at $p=400$.
	
	The largest differences in statistical performance occur under Scenario~3 (High-Leverage Outliers).
	As Table~\ref{tab:main_results} shows, the Standard MM-estimator exhibits increasing estimation
	error as $p$ and $\epsilon$ grow (e.g., at $p=400$ and $\epsilon=0.30$, MSE $\approx 3{,}592$),
	consistent with sensitivity to initialization under high-leverage contamination. The deterministic
	baseline remains accurate at lower dimensions, but at $p=400$ it also exhibits substantial errors as
	contamination increases (e.g., MSE $\approx 3{,}687$ at $\epsilon=0.30$), while incurring very large
	compute times. In contrast, ROBU remains stable across the considered contamination levels and
	achieves markedly smaller errors in the $p=400$ regime (e.g., MSE $\approx 161$ at $\epsilon=0.30$),
	while being substantially faster than both baselines.
	
	To better understand whether poor performance is driven by numerical nonconvergence or initialization quality, we tracked convergence diagnostics alongside True Positive (TP) and False Positive (FP) trimming rates. Across configurations, convergence warnings were rare: none occurred for ROBU, and only a few for the standard MM fit. False positive rates were near zero across methods. Moreover, in settings where ROBU achieved low MSE, it trimmed nearly all contaminated points, whereas the standard MM fit left a larger fraction untrimmed. This pattern is consistent with sensitivity to initialization under high-leverage contamination, rather than numerical instability of the IRWLS iterations.

	\subsection{Sensitivity to Block Dimensions} \label{subsec:sensitivity}
	
	The ROBU algorithm introduces a key hyperparameter: the number of blocks, $k$. As discussed in
	Section~\ref{sec:theory}, $k$ governs a trade-off between computational cost, robustness of the
	blockwise initialization, and the ability of each block update to capture multivariate structure.
	Smaller values of $k$ yield higher-dimensional block subproblems, requiring more elemental
	subsampling to obtain clean starting points under contamination. Larger values of $k$ reduce the
	block dimension $p/k$, but increase per-block overhead and may limit the ability of individual
	blocks to capture multivariate leverage structure. To assess this trade-off, we evaluated ROBU
	across the grid $k \in \{1, 5, 10, 20, 40, 100, 400\}$ on a dataset with $n = 1{,}500$, $p = 400$,
	and $\epsilon = 0.20$ high-leverage outliers.
	
\begin{figure}[h!]
	\centering
	\includegraphics[width=0.95\textwidth]{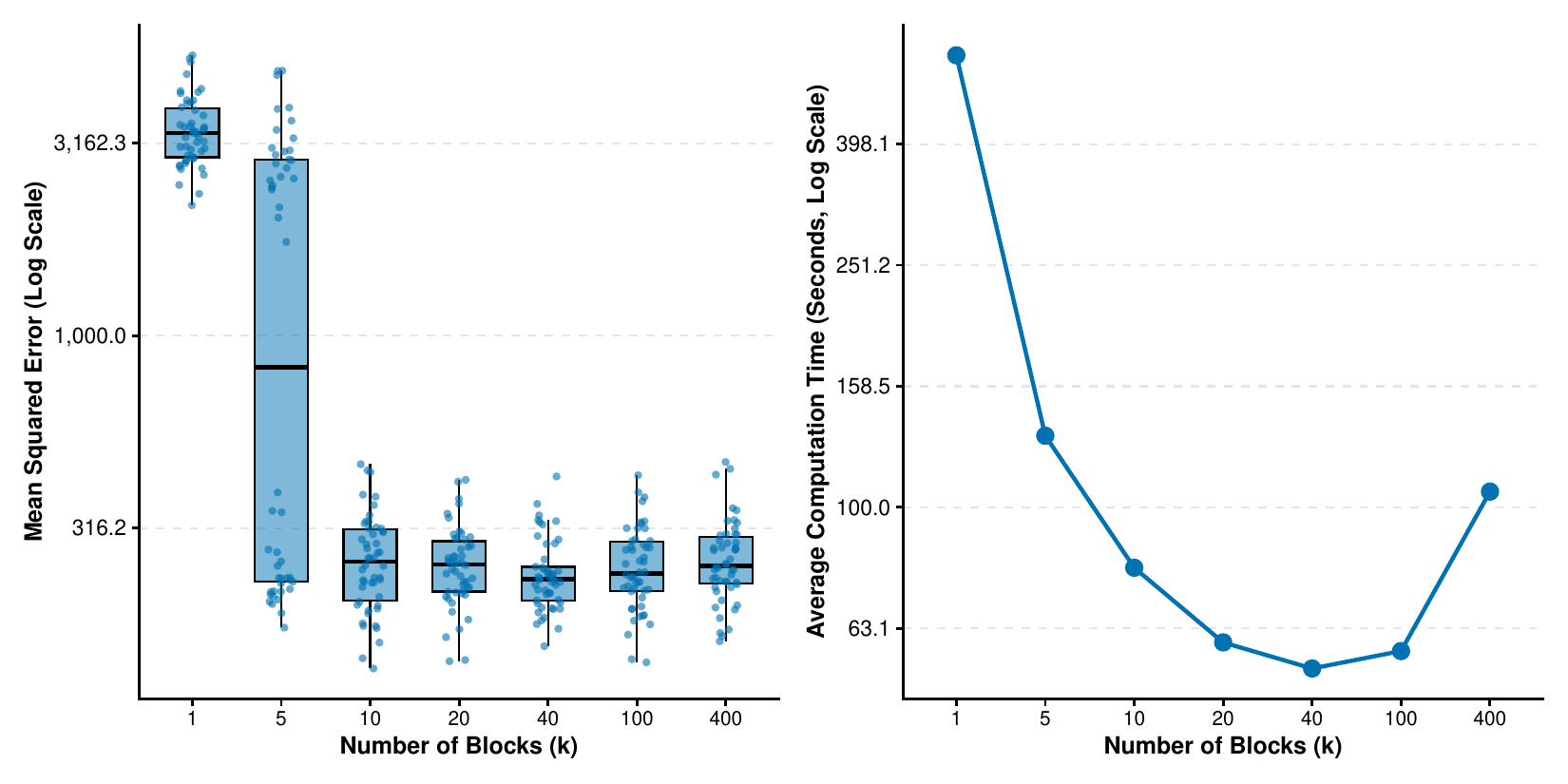}
	\caption{Sensitivity of the ROBU algorithm to the number of blocks ($k$) under 20\% high-leverage outliers ($n=\text{1,500}$, $p=400$). \textbf{(Left)} Boxplots of estimation accuracy (MSE) across 50 replications on a logarithmic scale. \textbf{(Right)} Average computation time (Seconds) across 50 replications on a logarithmic scale.}
	\label{fig:sensitivity}
	\end{figure}

	The results are shown in Figure~\ref{fig:sensitivity}. When $k$ is small (large block size), the
	blockwise initialization can be unreliable under a fixed resampling budget, leading to larger
	estimation error and increased variability. In particular, $k=1$ corresponds to the unblocked fit
	(i.e., the standard MM-estimator in the full $p$-dimensional problem), which performs poorly in
	this setting under the fixed resampling budget. As $k$ increases (smaller blocks), the MSE
	decreases substantially and the distributions become more concentrated. In this experiment, most of
	the improvement occurs once $k$ reaches moderate values (e.g., $k \in \{10,20,40\}$), after which
	the bulk of the MSE distributions are similar, although occasional outlying runs persist at
	$k=5$ and $k=10$.
	
	The right panel of Figure~\ref{fig:sensitivity} shows the corresponding computation times. For small $k$, runtimes are higher due to initialization on larger subproblems and costlier linear algebra. As $k$ increases, computation time decreases because these operations apply to smaller blocks. For very large $k$ (e.g., $k \ge 100$), runtime rises again due to per-block overhead and the sequential loop. Overall, moderate block sizes (10--20 variables per block) offered the best runtime-stability trade-off under high-leverage contamination, with the heuristic $k=\max(1,\lfloor p/10 \rfloor)$ performing well in this experiment.

	\section{Real Data Application: Proteogenomics} \label{sec:application}
	
	To illustrate ROBU in a real high-dimensional setting, we predict Estrogen Receptor Alpha (ER-$\alpha$) protein expression using RNA-sequencing data. Because mRNA--protein correlation is frequently attenuated by post-translational regulation and technical noise \citep{vogel2012insights}, this setting can exhibit heavy-tailed behavior and motivate robust regression. We utilized matched data from $n=882$ tumor samples in the TCGA Breast Invasive Carcinoma cohort \citep{weinstein2013cancer, curatedTCGAData_package}. To create a challenging moderate-to-high dimensional setting ($n \approx 3p$), we retained the 300 most variable genes, yielding a standardized predictor matrix $X \in \mathbb{R}^{882 \times 300}$. Since the true outlier structure of the raw data is unknown, we first fit the standard MM-estimator to the unmodified dataset to obtain a baseline reference coefficient vector, $\hat{\beta}_{\text{clean}}$.
	
	Over 50 replications, we assessed robustness by injecting 15\% high-leverage contamination ($m=132$ patients). For these patients, 30 randomly selected genes were perturbed with $\mathcal{N}(10,1)$ noise; their responses were generated to deviate substantially from the relationship captured by $\hat{\beta}_{\text{clean}}$. As in the simulation study, resampling-based procedures used the same \texttt{lmrob.control} settings (including \texttt{nResample}$=5{,}000$). We evaluated the estimators using computation time, squared deviation from the baseline coefficient vector ($\lVert \hat{\beta}_{\text{contam}} - \hat{\beta}_{\text{clean}} \rVert_2^2$), and True/False Positive detection rates computed from the final robust weights.
	
\begin{figure}[h!]
	\centering
	\includegraphics[width=0.95\textwidth]{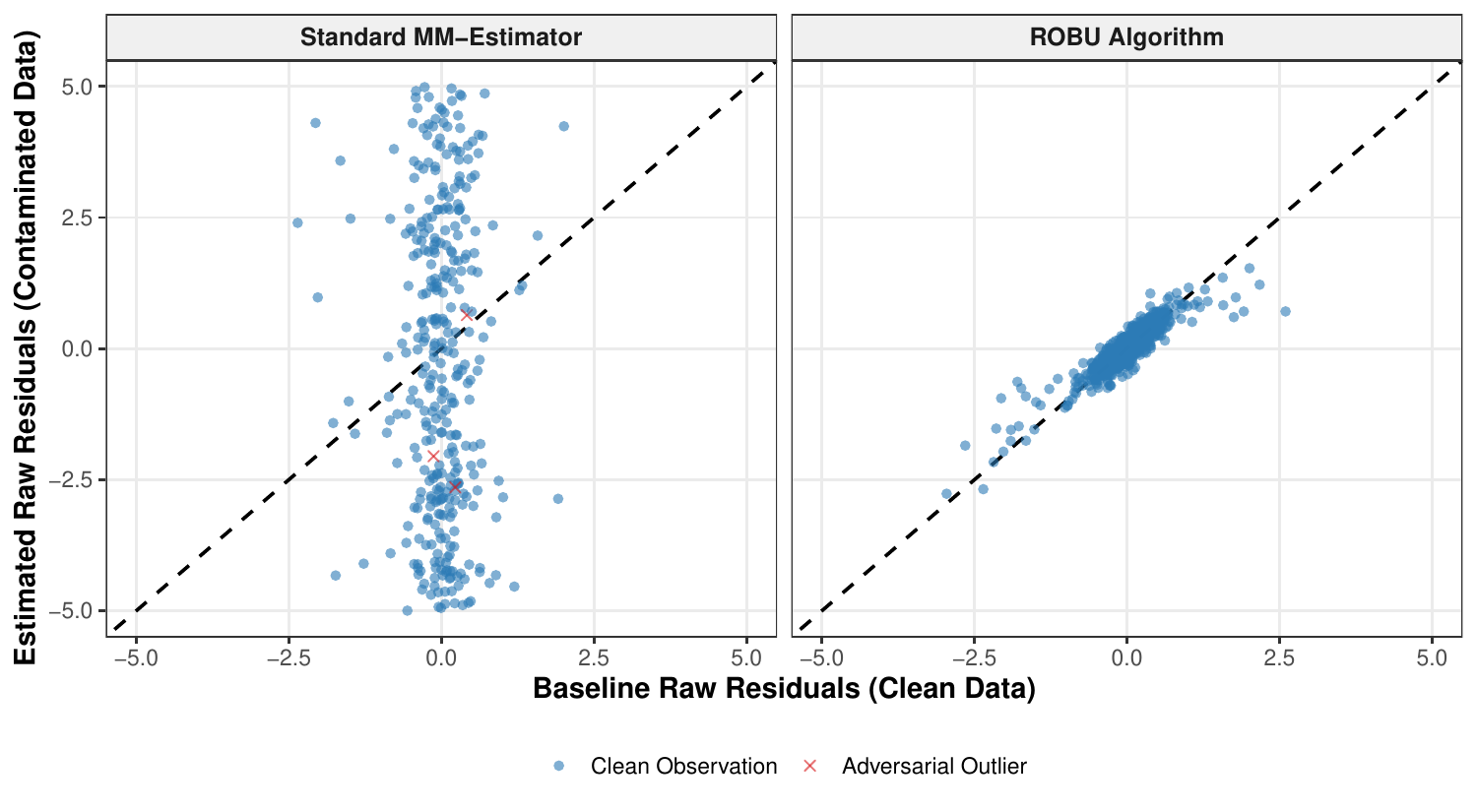}
	\caption{Raw regression residuals for the TCGA proteogenomics dataset. Contaminated model residuals ($y$-axis) are plotted against the clean baseline residuals ($x$-axis) for the Standard MM-estimator \textbf{(Left)} and ROBU \textbf{(Right)}. The dashed line ($y=x$) indicates identical residual values between the fits. Axis limits are truncated to $[-5, 5]$.}
	\label{fig:real_data_plot}
	\end{figure}

	Table~\ref{tab:real_data} summarizes the results. In this contamination experiment, the standard and deterministic MM fits yield substantially larger deviations from the baseline (MSE $> 2{,}000$) and leave a nontrivial fraction of contaminated observations untrimmed. In contrast, ROBU yields a coefficient vector much closer to the baseline (MSE $=0.5$) while trimming a larger fraction of injected high-leverage points, and does so in substantially less computation time ($\sim$35 seconds versus several minutes for standard MM). To visualize these differences, Figure~\ref{fig:real_data_plot} plots raw residuals from the contaminated fits against baseline residuals from the unmodified dataset.
	
	\begin{table}[h!]
	\centering
	\caption{Average estimator performance on the TCGA proteogenomics dataset across 50 replications with 15\% high-leverage contamination. True Positive (TP) and False Positive (FP) rates are percentages; computation time is in seconds.}
	\label{tab:real_data}
	\renewcommand{\arraystretch}{1.2}
	\begin{tabular}{l c rrrr}
		\toprule
		\textbf{Method} & & \textbf{MSE} & \textbf{TP Rate} & \textbf{FP Rate} & \textbf{Time} \\ 
		\midrule
		Std MM & & 2,215.2 & 87.8\% & 0.2\% & 273.0 \\ 
		Det MM & & 2,014.1 & 89.3\% & 1.0\% & 130.4 \\ 
		ROBU   & & 0.5     & 100.0\%& 0.0\% & 33.8 \\ 
		\bottomrule
	\end{tabular}
	\end{table}
	
	In the left panel (Standard MM) of Figure~\ref{fig:real_data_plot}, many observations deviate
	substantially from the identity line ($y=x$), indicating a fit that differs markedly from the
	baseline and leaves a noticeable fraction of the injected high-leverage points untrimmed. In
	contrast, the right panel (ROBU) shows much closer agreement between baseline and contaminated
	residuals for most observations, with contaminated points standing out via large residual
	discrepancies. This pattern is consistent with Section~\ref{sec:invariance}: because residuals are
	preserved under the QR reparameterization, blockwise initialization in the orthogonal basis does
	not change which observations appear atypical for a given fit.

	\section{Conclusion} \label{sec:conclusion}
	
	High-breakdown robust regression methods are indispensable for analyzing noisy, heavy-tailed data, but their reliance on random resampling creates  computational and statistical bottlenecks in high dimensions. In this paper, we proposed the Robust Orthogonal Block Updates (ROBU) algorithm as a scalable, theoretically sound solution for computing MM-estimators when the number of predictors $p$ is large but less than the sample size $n$.
	
	By decoupling the predictor space using a standard QR decomposition, ROBU allows for independent block-coordinate descent, completely bypassing the oscillatory behavior of naive block updates on correlated features. Furthermore, we showed that this orthogonal projection preserves the exact affine residuals and robust observation weights, allowing the block-wise estimators to safely identify and trim rotated leverage points without requiring a slow, computationally intensive robust orthogonalization step.
	
	ROBU's primary contribution is improving the reliability of subsampling-based initialization for MM-estimation in moderate-to-large dimensions. In these settings, standard unblocked implementations may fail to draw an outlier-free elemental subset within a fixed resampling budget, leading to highly biased fits under high-leverage contamination. ROBU addresses this by bounding the dimension of the subsampling subproblems via blockwise updates in an orthogonal basis. Our theoretical and empirical results demonstrate that this strategy substantially reduces computation time and improves robustness. The proteogenomics application further illustrates this, showing that ROBU yields fits closer to a clean baseline and trims a larger fraction of injected high-leverage points than competing methods.
	
	Future extensions of this work could explore integrating scalable block-update frameworks with robust penalized methods, such as the Penalized Elastic Net S-Estimator (PENSE) \citep{cohen2019robust} and its adaptive extensions \citep{kepplinger2023robust}. While orthogonal rotations fundamentally compromise the coordinate-wise separability of sparsity-inducing $\mathcal{L}_1$ penalties, leveraging operator-splitting techniques (e.g., the Alternating Direction Method of Multipliers) to isolate the robust loss could yield significant computational accelerations for robust variable selection and prediction in the ultra-high dimensional $p > n$ regime.
	
	
	\section*{Statements and Declarations}
	
	\begin{description}
	\item[Funding:] The authors did not receive support from any organization for the submitted work. 
	
	\item[Competing Interests:] The authors declare that they have no relevant financial or non-financial competing interests to report.
	
	\item[Data and Code Availability:] The complete \texttt{R} implementation of the ROBU algorithm, alongside all scripts and datasets necessary to fully reproduce the simulation studies and the proteogenomics data application, are publicly available on GitHub at \url{https://github.com/AnthonyChristidis/ROBU}.
	
	\item[Author Contributions:] All authors contributed to the study conception and design. Material preparation, data collection, and analysis were performed by Anthony Christidis and Matías Salibián-Barrera. The first draft of the manuscript was written by Anthony Christidis, and all authors commented on previous versions of the manuscript. All authors read and approved the final manuscript.
	\end{description}

	\appendix
	
	\titleformat{\section}{\Large\bfseries}{Appendix \thesection:}{0.5em}{}
	
	\section{Effect of Distributing the Blockwise Subsampling Budget}
	\label{sec:appendixA}
	
	Section~\ref{sec:theory} derives the number of elemental sub-samples required to obtain, with high probability,
	at least one clean starting point, and shows that this resampling budget can be distributed across
	the $k$ blockwise subproblems. In the main simulation study in Section~\ref{sec:sims}, we used a conservative
	configuration in which the same \texttt{lmrob.control} resampling budget used for the full
	$p$-dimensional problem was also passed to each blockwise fit in ROBU's initialization (Step~3 of
	the ROBU algorithm).
	
	Here we evaluate an alternative configuration in which the blockwise resampling budget is
	distributed across blocks. We consider the configuration $n=1{,}500$, $p=400$, and high-leverage
	contamination with $\epsilon \in \{0.20,0.30\}$. With the default heuristic $k=\lfloor p/10 \rfloor$,
	ROBU uses $k=40$ blocks of size $p/k=10$.
	
	We compare the following two ROBU variants (each evaluated over 50 independent replications):
	\begin{enumerate}
		\item \textbf{Full per-block budget.} The blockwise initialization uses \texttt{nResample}$=5{,}000$
		in each block, i.e., $40 \times 5{,}000 = 200{,}000$ total elemental sub-samples across the
		initialization stage.
		\item \textbf{Distributed per-block budget.} The blockwise initialization uses
		\texttt{nResample}$=\lceil 5{,}000/k \rceil = 125$ per block, so that the total number of
		elemental sub-samples across all blocks is approximately $5{,}000$.
	\end{enumerate}
	In both cases, the final global MM stage (Steps~4--6 of the ROBU algorithm) is held fixed; only the
	blockwise subsampling budget used in the initialization differs.
	
\begin{figure}[h!]
		\centering
		\includegraphics[width=0.65\textwidth]{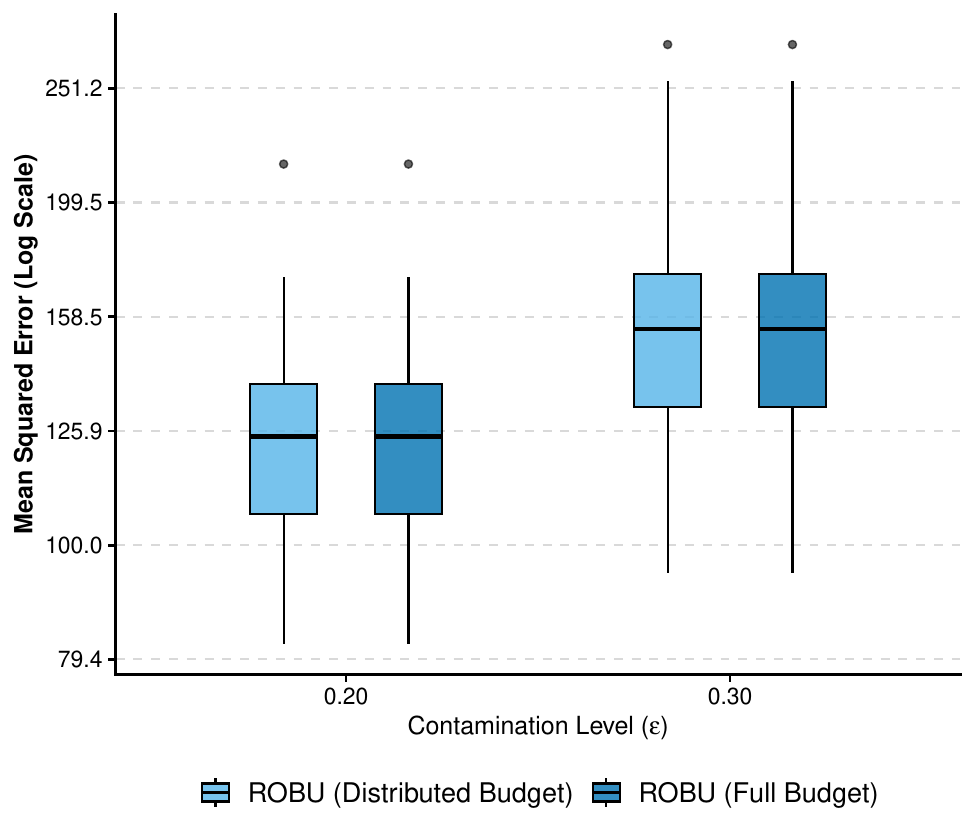}
		\caption{ROBU estimation error (MSE) under full (\texttt{nResample}=5{,}000) versus distributed (\texttt{nResample}=$\lceil 5{,}000/k\rceil$) per-block resampling budgets. Results shown for $n=1{,}500$, $p=400$ ($k=40$ blocks), and high-leverage contamination over 50 replications.}
		\label{fig:robu_budget_comparison}
	\end{figure}

	Figure~\ref{fig:robu_budget_comparison} shows the distribution of estimation error (MSE) for both configurations at each contamination level, while Table~\ref{tab:supp_budget_times} reports the corresponding average MSE and computation times. The results demonstrate that distributing the budget yields essentially identical estimation accuracy while reducing computation time by more than half.
	
	\begin{table}[h!]
		\centering
		\caption{Average Mean Squared Error (MSE) and Computation Time (Seconds) for the ROBU blockwise subsampling budget comparison across 50 replications ($n=1{,}500$, $p=400$). \vspace{0.5em}}
		\label{tab:supp_budget_times}
		\renewcommand{\arraystretch}{1.2}
		\begin{tabular}{l rr c rr}
			\toprule
			& \multicolumn{2}{c}{$\boldsymbol{\epsilon = 0.20}$} & & \multicolumn{2}{c}{$\boldsymbol{\epsilon = 0.30}$} \\
			\cmidrule(lr){2-3} \cmidrule(lr){5-6}
			\textbf{Configuration} & \textbf{MSE} & \textbf{Time} & & \textbf{MSE} & \textbf{Time} \\
			\cmidrule(lr){1-1} \cmidrule(lr){2-3} \cmidrule(lr){5-6}
			ROBU (Distributed Budget) & 123.7 & 65.9  & & 161.8 & 34.0  \\
			ROBU (Full Budget)        & 123.7 & 139.4 & & 161.8 & 108.3 \\
			\bottomrule
		\end{tabular}
	\end{table}

	\section{Extended Simulation Results} \label{sec:appendixB}
	
	In Section~\ref{sec:sims}, the primary table reports computation time and estimation
	accuracy under clean data ($\epsilon=0$) and high-leverage outliers (Scenario~3), which are the most
	challenging settings for robust initialization. To complement those results, Table~\ref{tab:supp_vertical}
	reports performance under Scenario~2 (Vertical Outliers), where contamination affects only the
	responses and the predictor matrix $\mathbf{X}$ is unchanged. Because vertical outliers do not
	alter the leverage structure of $\mathbf{X}$, they are often easier to downweight during the
	refinement stage of robust procedures. These results therefore help separate failures driven by
	leverage contamination from those driven purely by response contamination, while also providing
	additional context on computational scaling at $n=1{,}500$ as $p$ increases.
	
	In the vertical-outlier setting, contaminated observations affect only the responses $y_i$ while
	leaving the predictor matrix $\mathbf{X}$ unchanged. In many moderate-dimensional settings, robust
	MM procedures can downweight such observations effectively during the refinement (M-estimation)
	stage. Table~\ref{tab:supp_vertical} shows that this behavior persists in our experiments for the
	resampling-based MM-estimator and for ROBU, which yield similar estimation errors across
	dimensions, with ROBU substantially faster. In contrast, the deterministic baseline
	(\texttt{lmrobdetMM}) can become unstable in the largest configuration: at $p=400$ and
	$\epsilon=0.30$ it exhibits a very large estimation error (MSE $\approx 2{,}053{,}868.8$), despite
	performing similarly to the other robust procedures at smaller contamination levels.
	
	Across the configurations we examined, convergence warnings were rare for the robust procedures
	(see the diagnostics reported in the main simulations). In particular, we did not observe
	systematic convergence issues for ROBU, suggesting that the differences in estimation error are
	primarily driven by the resulting robust weights and effective initialization, rather than numerical
	nonconvergence of the IRWLS refinement steps. Table~\ref{tab:supp_vertical} also illustrates the
	computational advantage of ROBU in this vertical-outlier regime, especially at higher dimensions.
	Additional results on distributing the blockwise subsampling budget (Section~\ref{sec:theory}) are provided in
	Appendix~\ref{sec:appendixA}.

	\begin{table}[h!]
		\centering
		\caption{Average Mean Squared Error (MSE) and Computation Time (Seconds) under Scenario~2 (Vertical Outliers) across 50 replications ($n=1{,}500$). \vspace{0.5em}}
		\label{tab:supp_vertical}
		\renewcommand{\arraystretch}{1.2}
		\setlength{\tabcolsep}{4pt}
		\begin{tabular}{ll rr c rr c rr c rr}
			\toprule
			& & \multicolumn{2}{c}{$\boldsymbol{\epsilon = 0.10}$} & & \multicolumn{2}{c}{$\boldsymbol{\epsilon = 0.20}$} & & \multicolumn{2}{c}{$\boldsymbol{\epsilon = 0.30}$} \\
			\cmidrule(lr){1-2} \cmidrule(lr){3-4} \cmidrule(lr){6-7} \cmidrule(lr){9-10}
			$\boldsymbol{p}$ & \textbf{Method} & \multicolumn{1}{c}{\textbf{MSE}} & \multicolumn{1}{c}{\textbf{Time}} & & \multicolumn{1}{c}{\textbf{MSE}} & \multicolumn{1}{c}{\textbf{Time}} & & \multicolumn{1}{c}{\textbf{MSE}} & \multicolumn{1}{c}{\textbf{Time}} \\
			\cmidrule(lr){1-2} \cmidrule(lr){3-4} \cmidrule(lr){6-7} \cmidrule(lr){9-10}
			
			100 & OLS     & 17,221.3 & 0.0   & & 33,332.4 & 0.0   & & 48,606.5   & 0.0 \\
			& Std MM  & 4.8      & 37.2  & & 5.4      & 33.9  & & 5.9        & 33.2 \\
			& Det MM  & 4.8      & 55.9  & & 5.4      & 53.3  & & 5.9        & 51.8 \\
			& ROBU    & 4.8      & 11.6  & & 5.4      & 10.2  & & 5.9        & 8.8 \\
			
			\addlinespace[0.7em]
			
			200 & OLS     & 76,498.9 & 0.1   & & 148,889.1 & 0.1  & & 222,767.3  & 0.1 \\
			& Std MM  & 22.2     & 188.9 & & 25.1      & 162.0 & & 28.9       & 149.0 \\
			& Det MM  & 22.2     & 403.5 & & 24.9      & 387.9 & & 28.8       & 297.3 \\
			& ROBU    & 22.2     & 31.4  & & 25.1      & 30.7  & & 28.9       & 21.1 \\
			
			\addlinespace[0.7em]
			
			400 & OLS     & 375,314.9 & 0.1  & & 752,395.9 & 0.1  & & 1,228,635.6 & 0.1 \\
			& Std MM  & 111.5    & 855.0 & & 131.2     & 854.1 & & 172.5       & 809.7 \\
			& Det MM  & 113.7    & 1,891.8& & 131.2     & 1,850.0& & 2,053,868.8 & 1,915.3 \\
			& ROBU    & 111.3    & 86.0  & & 131.2     & 84.9  & & 172.5       & 67.5 \\
			\bottomrule
		\end{tabular}
	\end{table}

	\section{Simulation and Software Details} \label{sec:appendixC}
	
	To support reproducibility and facilitate comparison across methods, all simulations and real data analyses were executed using the same software environment and closely matched algorithmic tuning parameters.
	
	\subsection{Software and Hardware Environment}
	All computations were performed in \texttt{R} version 4.4.2. Monte Carlo simulations and real data analyses were executed on the O2 High-Performance Computing Cluster at Harvard Medical School. To reduce differences in implicit multithreading affecting timing results, each task was restricted to a single computational thread (via OpenBLAS/MKL environment variables) and allocated 8--12~GB of RAM.
	
	The standard resampling-based MM-estimator was computed using the \texttt{lmrob} function from the \texttt{robustbase} package (version 0.99-6) \citep{maechler2023robustbase}. The deterministic MM-estimator was computed using the \texttt{lmrobdetMM} function from the \texttt{RobStatTM} package (version 1.0.11) \citep{yohai2024robstattm}. In the ROBU implementation, robust M-scale estimates required during the S-initialization and refinement steps were computed using the \texttt{mscale} routine from \texttt{RobStatTM}. Multivariate normal data generation used the \texttt{mvnfast} package (version 0.2.8) \citep{fasiolo2014mvnfast}.
	
	\subsection{Algorithmic Tuning Parameters}
	The \texttt{robustbase} implementation of the MM-estimator relies on a highly complex set of control parameters that govern the initialization, sub-sampling, and convergence criteria. To isolate the effect of the orthogonal block-coordinate descent architecture, the ROBU algorithm was explicitly configured to inherit the exact same control parameters as the baseline \texttt{lmrob} function.
	
	The control list, \texttt{my.control}, was defined using \texttt{lmrob.control()} with the following critical modifications to enforce the fixed computational budget:

	\begin{itemize}
		\item \texttt{method = "MM"}: Both algorithms were restricted to the classic MM-estimator chain (initial S-estimate followed by an M-estimate for efficiency) rather than the default SMDM chain. This ensures the benchmark evaluates the core Fast-S optimization bottleneck without the confounding effects of additional Design-based scale estimations.
		\item \texttt{nResample}, \texttt{k.max}, \texttt{k.m\_s}, \texttt{max.it}: These parameters control the subsampling and refinement budget of the resampling-based procedures. Relative to the default \texttt{robustbase} settings, we increased them to provide a larger budget in higher dimensions. Specifically, for the $p=400$ regime, we set \texttt{nResample}$=\texttt{5000}$ and \texttt{k.max}$=\texttt{5000}$, and for all configurations we used \texttt{k.m\_s}$=\texttt{2000}$ and \texttt{max.it}$=\texttt{1000}$. Because ROBU inherits this same \texttt{lmrob.control} list, the blockwise initialization in the main simulations applies this resampling budget within each block (e.g., 5,000 sub-samples per block when $p=400$). Appendix~\ref{sec:appendixA} additionally evaluates a configuration in which the blockwise resampling budget is distributed across blocks.
		Due to the $\mathcal{O}(p^3 / k^3)$ reduction in matrix inversion complexity, ROBU can evaluate many more blockwise candidates within a comparable wall-clock time budget.
		\item \texttt{fast.s.large.n = Inf}: By default, the \texttt{robustbase} package employs a heuristic that switches away from the standard Fast-S elemental subsampling algorithm when the sample size $n$ exceeds a certain threshold (typically $n > \text{2,000}$). To maintain algorithmic consistency across all experimental regimes and mathematically enforce the probabilistic subsampling framework described in Section~\ref{sec:theory}, this parameter was set to infinity. This forces the standard MM-estimator to rely exclusively on the Fast-S algorithm.
		\item \texttt{refine.tol = 1e-5}: The convergence tolerance for the IRWLS refinement steps was slightly relaxed from the default \texttt{1e-7} to \texttt{1e-5}. Because the ROBU algorithm fits regressions on perfectly orthogonalized sub-matrices ($\mathbf{Q}_j$) without an intercept term, the gradient of the loss function frequently flattens rapidly. This relaxed tolerance prevents the internal Fast-S routines from prematurely throwing ``refinements did not converge'' warnings due to orthogonal geometry, without materially altering the final Mean Squared Error (MSE) of the estimates.
		\item \textbf{Breakdown Point ($\delta$):} The tuning constants for the Tukey bisquare loss functions ($\rho_0$ and $\rho_1$) were set to their default values, guaranteeing an asymptotic breakdown point of 50\% and an asymptotic efficiency of 95\% under Gaussian errors for both algorithms.
	\end{itemize}
	
	\subsection{Block Size Selection ($k$)}
	For all simulations and the real-data application in the main manuscript, the number of blocks in ROBU was chosen as a function of the predictor dimension $p$. Specifically, we used the default heuristic described in Section~\ref{subsec:blocks}, $k = \max\!\bigl(1, \lfloor p/10 \rfloor \bigr)$, so that the typical block size is on the order of 10 variables. The sensitivity analysis in Section~\ref{subsec:sensitivity} illustrates the effect of varying $k$ in a representative high-leverage contamination setting.

	\bibliographystyle{apalike}  
	\bibliography{ROBU} 
	
\end{document}